\documentclass[10pt,conference,letterpaper]{IEEEtran}
\usepackage{cite}
\usepackage{graphicx,color,epsfig,rotating}
\usepackage{amsfonts,amsmath,amssymb}
\usepackage{algorithm,algorithmic}
\usepackage{subfigure}
\long\def\comment#1{}

\newfont{\bbb}{msbm10 scaled 700}

\newfont{\bb}{msbm10 scaled 1100}
\newcommand{\CC}{\mbox{\bb C}}

\newcommand{\EE}{\mbox{\bb E}}

\newcommand{\nv}{{\bf n}}

\newcommand{\rv}{{\bf r}}
\newcommand{\sv}{{\bf s}}

\newcommand{\wv}{{\bf w}}

\newcommand{\xv}{{\bf x}}
\newcommand{\yv}{{\bf y}}
\newcommand{\zv}{{\bf z}}
\newcommand{\zerov}{{\bf 0}}

\newcommand{\Hm}{{\bf H}}
\newcommand{\Id}{{\bf I}}

\newcommand{\Qm}{{\bf Q}}
\newcommand{\Rm}{{\bf R}}

\newcommand{\Wm}{{\bf W}}

\newcommand{\Cc}{{\cal C}}

\newcommand{\Gc}{{\cal G}}

\newcommand{\Kc}{{\cal K}}
\newcommand{\Lc}{{\cal L}}
\newcommand{\Mc}{{\cal M}}
\newcommand{\Nc}{{\cal N}}

\newcommand{\Rc}{{\cal R}}

\newcommand{\muv}{\hbox{\boldmath$\mu$}}

\newcommand{\Thetam}{\hbox{\boldmath$\Theta$}}

\newcommand{\diag}{{\hbox{diag}}}

\newcommand{\trace}{{\hbox{tr}}}

\newcommand{\herm}{{\sf H}}

\newcommand{\transp}{{\sf T}}

\newcommand{\MMSE}{{\sf mmse}}

\begin{document}

\title{Multi-cell MIMO Downlink with Fairness Criteria: the Large System Limit}
\author{\IEEEauthorblockN{Hoon Huh\IEEEauthorrefmark{1},
Giuseppe Caire\IEEEauthorrefmark{1},
Sung-Hyun Moon\IEEEauthorrefmark{2}, and Inkyu Lee\IEEEauthorrefmark{2}}
\IEEEauthorblockA{\IEEEauthorrefmark{1}University of Southern California, Los Angeles,
CA, USA, Email: hhuh, caire@usc.edu}
\IEEEauthorblockA{\IEEEauthorrefmark{2}Korea University, Seoul, Korea, Email: shmoon,
inkyu@korea.ac.kr}
}

\maketitle

\begin{abstract}
We consider the downlink of a cellular network with multiple cells and multi-antenna base
stations under arbitrary inter-cell cooperation, realistic distance-dependent pathloss,
and general ``fairness'' requirements. Beyond Monte Carlo simulation, no efficient computation
method to evaluate the ergodic throughput of such systems has been provided so far.
We propose an analytic method based on the combination of the large random matrix theory with
Lagrangian optimization. The proposed method is computationally much more efficient than Monte Carlo
simulation and provides a very accurate approximation (almost indistinguishable) for the actual
finite-dimensional case, even for of a small number of users and base station antennas.
Numerical examples include linear 2-cell and planar three-sectored 7-cell layouts, with
no inter-cell cooperation, sector cooperation, and full cooperation.
\end{abstract}

\section{Introduction} \label{sec:intro}

Multiuser MIMO (MU-MIMO) technology is expected to play a key role in future wireless cellular
networks. The MIMO Gaussian broadcast channel serves as an information theoretic
model for the MU-MIMO downlink and its capacity was studied and characterized in
\cite{Caire-Shamai-TIT03, Goldsmith-JSAC03, Weingarten-Steinberg-Shamai-TIT06}.
In a multi-cell scenario, depending on the level of base station (BS) cooperation, we are in the
presence of a MIMO broadcast and interference channel problem, which is not yet fully understood
in the information theoretic sense. In this work, we restrict our attention to the case where
clusters of cooperating BSs act as a single distributed MIMO transmitter and where interference
from other clusters of BSs is treated as noise. Our framework encompasses arbitrary cooperation
clusters.

Multi-cell systems under full and partial cooperation have been widely studied
under the so-called ``Wyner model'' \cite{Wyner-TIT94, Shamai-Wyner-TIT97, Somekh-Shamai-TIT00,
Sanderovich-TIT09} which assumes a very simplified and rather unrealistic model
in regard to the pathloss and makes the system essentially identical with respect to any given user.
As a matter of fact, users in different locations under the cellular coverage are subject to
distance-dependent pathloss that may have more than 30 dB dynamic range \cite{Wimax-eval06}.
It follows that in these realistic propagation conditions the channels from a BS to users
are not identically distributed at all. Furthermore, studying the downlink sum-capacity
(or achievable sum-throughput, under some suboptimal scheme) is not a very meaningful approach
from a system performance viewpoint: rate and power optimization would lead to the solution of
letting a BS transmit only to users very close to the BS, while leaving users at the cell
edges to starve.
In order to avoid this problem, {\em downlink scheduling} has been widely studied, which achieves
a good balance between network efficiency and ``fairness'' (see for example
\cite{Viswanath-Tse-Laroia-TIT02, Georgiadis-Neely-Tassiulas-04, ShiraniMehr-Caire-Neely-submit09}
and refs. therein). The goal of fairness scheduling is to make the system operate at a rate point
in its ergodic achievable rate region such that a suitable {\em network utility function}
is maximized \cite{Mo-TNET00}. Taking into account realistic pathloss models,
multi-cell systems with arbitrary inter-cell cooperation, and fairness requirements
makes system analysis extremely complicated. In fact, so far, the system performance of such systems
has been evaluated only through extensive and computationally very intensive Monte Carlo simulation.

In this work, we consider a particular ``large system limit'' that is studied with known
results from the large random matrix theory \cite{Tulino-TIT05, Aktas-TIT06, Moon-Caire-ICC10}
and combine it with Lagrangian optimization in order to obtain a numerical ``almost'' closed-form
analysis tool that incorporates all the above system aspects. The proposed method is much more
efficient than Monte Carlo simulation and, somehow surprisingly, provides results that match
very closely the performance of finite-dimensional systems, even for very small dimension.

\section{System setup} \label{sec:setup}

Consider $M$ BSs with $\gamma N$ antennas each, and $KN$ single-antenna user terminals
distributed in the cellular coverage region. Users are divided into $K$ co-located ``user groups''
of equal size $N$. Users in the same group are statistically equivalent: they experience the same
pathloss from all BSs and their small-scale fading channel coefficients are iid.
\footnote{In practice, co-located users are separated by a sufficient number of wavelength
such that they undergo iid small-scale fading. However, the wavelength is sufficiently small so that
they all have essentially the same distance-dependent pathloss.}
The received signal vector $\yv_k = [y_{k,1}\cdots y_{k,N}]^\transp \in \CC^N$ for user group $k$
is given by
\begin{equation} \label{eq:ch_model}
\yv_k = \sum_{m=1}^{M} \alpha_{m,k} \Hm_{m,k}^\herm \xv_m + \nv_k
\end{equation}
where $\alpha_{m,k}$ and $\Hm_{m,k}$ denote the the distance dependent pathloss and
$\gamma N \times N$ small-scale channel fading matrix from BS $m$ to user group $k$,
respectively, $\xv_m = [x_{m,1}\cdots x_{m,\gamma N}]^\transp \in\CC^{\gamma N}$ is the transmitted
signal vector of BS $m$ subject to the power constraint $\trace \left( {\rm Cov}(\xv_m) \right) \leq P$,
and $\nv_k=[n_{k,1}\cdots n_{k,N}]^\transp \in \CC^N$ denotes the AWGN at the user terminals.
The elements of $\nv_k$ and of $\Hm_{m,k}$ are iid $\sim \Cc\Nc(0,1)$.

We define the BS partition $\{\Mc_1,\cdots,\Mc_L\}$ of set $\{1,\cdots, M\}$ and the
corresponding user group partition $\{\Kc_1,\cdots, \Kc_L\}$ of set $\{1,\cdots, K\}$, where
$L$ is the number of cooperation clusters. Cooperation cluster $\ell$ is formed by the BSs
in $\Mc_\ell$, acting effectively as a distributed network MIMO transmitter, and the user groups
in $\Kc_\ell$. We assume that the BSs in each cluster have perfect channel state information (CSI)
for all users in the same cluster and {\em statistical information} (i.e. known distributions
but not the instantaneous values) relative to channels from BSs in other clusters.
The impact of non-perfect CSI at the transmitter and the required overhead for CSI estimation
for inter-cell cooperation is addressed in \cite{Huh-Tulino-Caire-CISS10}.
We restrict to the symmetric cellular structure for clusters such that each cluster
has the same number of BSs and the same distribution of user groups with respect
to all the BSs in the cluster. Specific examples of such a structure will be given in Section
\ref{sec:result}. Under these assumption, owing to the symmetric nature of clusters,
the sum transmit power of each BS is equal to $P$ when we consider the cluster-wise sum-power
constraint $|\Mc_\ell|P$ and it is satisfied with an equality.
Then, the interference plus noise variance at any user terminal in group $k \in \Kc_\ell$ is obtained as
\begin{equation} \label{eq:ici-pow}
\sigma_k^2 = 1 + \sum_{m \notin \Mc_\ell} \alpha_{m,k}^2 P
\end{equation}
and, from the viewpoint of cluster $\ell$, the system is equivalent to a single-cell
MU-MIMO downlink given by
\begin{equation} \label{eq:single-bc}
\yv_\ell = \Hm_\ell^\herm \xv_\ell + \zv_\ell
\end{equation}
where $\yv_\ell \in \CC^{|\Kc_\ell| N}$ is the received signal vector formed by
$\{\yv_k : k \in \Kc_\ell\}$, $\zv_\ell \in \CC^{|\Kc_\ell| N}$ is a Gaussian interference plus
noise vector with independent elements and block-diagonal covariance matrix
$\EE[\zv_\ell \zv_\ell^\herm] = \diag( \sigma_k^2 \Id_{N} : k \in \Kc_\ell)$, and
$\Hm_\ell \in \CC^{\gamma |\Mc_\ell| N \times |\Kc_\ell| N}$ is formed as
\begin{align} \label{eq:H}
\Hm_\ell =\!\left[\!
 \begin{array}{ccc}
  \!\alpha_{m_1,k_1} \Hm_{m_1,k_1} & \!\!\cdots\!\! &
   \alpha_{m_1,k_{|\Kc_\ell|}}\Hm_{m_1,k_{|\Kc_\ell|}}\!  \\
  \!\vdots                         &            & \vdots\!             \\
  \!\alpha_{m_{|\Mc_\ell|},k_1}\Hm_{m_{|\Mc_\ell|},k_1} & \!\!\cdots\!\! &
   \alpha_{m_{|\Mc_\ell|},k_{|\Kc_\ell|}}\Hm_{m_{|\Mc_\ell|},k_{|\Kc_\ell|}}\!
 \end{array}\!\right]\!.
\end{align}

For the sake of system optimization, we introduce the ``dual uplink'' channel model
\cite{Goldsmith-JSAC03, Jindal-TIT03, Jindal-TIT05} corresponding to (\ref{eq:single-bc}) given by
\begin{equation} \label{eq:mac_model}
\rv = \widetilde{\Hm} \sv + \wv
\end{equation}
where we define the transmit signal $\sv \in \CC^{A N}$ with diagonal covariance matrix $\Qm$
subject to $\trace(\Qm) \leq Q$, the AWGN vector $\wv \sim \Cc\Nc(\zerov, \Id_{B N})$,
and the channel matrix $\widetilde{\Hm} \in \CC^{B N \times A N}$ formed by blocks
$\beta_{m,k} \Hm_{m,k}$ in the same way as in (\ref{eq:H}), and where, for notation simplicity,
we let $A = |\Kc_\ell|$, $B = \gamma |\Mc_\ell|$, $Q = |\Mc_\ell|P$,
and $\beta_{m,k} = \frac{\alpha_{m,k}}{\sigma_k}$.


We assume that the pathloss coefficients are fixed while the small-scale fading coefficients
change in time according to some ergodic process with the assigned first-order iid $\Cc\Nc(0,1)$
distribution. This is representative of a typical situation where the distance between BSs and
users changes significantly over a time-scale of the order of tens of seconds, whereas the small-scale
fading decorrelates completely within a few milliseconds \cite{Tse-05}. In this regime, the goal
of downlink scheduling is to maximize a suitable strictly increasing and concave network utility
function $g(\cdot)$ of ergodic user rates \cite{Mo-TNET00}. By symmetry, all users in the same
group should achieve the same ergodic rate. However, users in different groups may operate at
different rates, depending on the network utility function $g(\cdot)$. For
$\Rm = [R_1,R_2,\cdots,R_A]$ where $R_k$ is the ergodic group rate given by the mean user rate
of user group $k$, the fairness scheduling problem is formulated as
\begin{align} \label{eq:sche}
\mbox{maximize} \;\;\; & g(\Rm) \nonumber \\
\mbox{subject to} \;\;\; & \Rm \in \Rc_{\rm erg}(Q)
\end{align}
where $\Rc_{\rm erg}(Q)$ is the region of ergodic achievable group rates for system
(\ref{eq:mac_model}) (equivalently, for cluster $\ell$ with channel model defined in
(\ref{eq:single-bc})) for given sum-power constraint $Q$.


\vspace{2mm}
\section{Weighted ergodic sum-rate maximization} \label{sec:wesrm}

In this section, we recall the computation algorithm of \cite{Moon-Caire-ICC10} to solve,
in the limit for $N \rightarrow \infty$, the weighted ergodic sum-rate maximization problem:
\begin{align} \label{eq:wsrm}
\mbox{maximize} \;\;\; & \sum_{k=1}^A W_k R_k \nonumber \\
\mbox{subject to} \;\;\; & \Rm \in \Rc_{\rm erg}(Q).
\end{align}
This is a fundamental building block for the computation of (\ref{eq:sche}). The exposition
in this section is necessarily brief and more details can be found in \cite{Moon-Caire-ICC10}
and references therein.

Let $\pi$ denote the permutation such that $W_{\pi_1} \leq \cdots \leq W_{\pi_A}$, and define
$\Thetam_{k:A} = \widetilde{\Hm}_{k:A} \Qm_{k:A} \widetilde{\Hm}_{k:A}^\herm$, where
$\widetilde{\Hm}_{k:A}$ and $\Qm_{k:A}$ denote the channel and the input covariance sub-matrices
restricted to user groups $\pi_k, \ldots, \pi_A$. By symmetry, we have that all users
in the same group $k$ are allocated the same (uplink) power $Q_k$. After dividing the channel coefficients
by $\sqrt{N}$, the power constraint for any $N$ is given by $\sum_{k=1}^A Q_k \leq Q$.
For the dual uplink channel (\ref{eq:mac_model}), the maximum weighted sum-rate is achieved by
a block successive interference cancelation decoding strategy that jointly decodes users
from group $\pi_1$ to $\pi_A$, and, after decoding each group, subtracts it from the received
signal. It follows that problem (\ref{eq:wsrm}) reduces to the maximization with respect
to $\Qm$ of the objective function
\begin{equation} \label{sum-rate-obj}
F_\Wm(\Qm) = \sum_{k=1}^A \Delta_k \frac{1}{N}\EE \left[ \log \left| \Id + \Thetam_{k:A}
 \right| \right]
\end{equation}
where $\Delta_k = W_{\pi_k} - W_{\pi_{k-1}}$ with $W_{\pi_0} \triangleq 0$. From
the \textit{Karush-Kuhn-Tucker} (KKT) conditions applied to the Lagrangian function
$\Lc(\Qm, \lambda) = F_\Wm(\Qm) - \lambda ( \trace (\Qm) - Q)$, we can eliminate the Lagrange
multiplier $\lambda$ by imposing that the power constraint holds with equality and then we obtain
\begin{equation} \label{eq:KKT-finite-dim}
Q_{\pi_k} = Q \frac{\sum_{i=1}^k \Delta_i \left( 1 - \EE \left[\MMSE^{(k)}_{i:A} \right] \right)}
{\sum_{j=1}^A  \sum_{i=1}^j \Delta_i \left( 1 - \EE \left[ \MMSE^{(j)}_{i:A} \right] \right) }
\end{equation}
where $\MMSE^{(k)}_{i:A} = \frac{1}{N} \trace ( \EE [ \Id - Q_{\pi_k}
\widetilde{\Hm}_{\pi_k}^\herm (\Id + \Thetam_{j:A})^{-1} \widetilde{\Hm}_{\pi_k} ] )$
for $k \geq i$ is the average {\em Minimum Mean Square Error} (MMSE)
for the input symbols of group $\pi_k$ when the symbols of groups $\pi_1, \ldots, \pi_{i-1}$ are
cancelled from the observation $\rv$ in (\ref{eq:mac_model}), and $\widetilde{\Hm}_{\pi_k}$
denotes the $BN \times N$ submatrix of $\widetilde{\Hm}$ corresponding to group $\pi_k$.

For finite $N$, the amount of calculation in order to evaluate the solution of
(\ref{eq:KKT-finite-dim}) 
is tremendous because $\EE [\MMSE^{(k)}_{i:A} ]$ must be computed by Monte Carlo simulation.
In the following, we consider the large system regime where we let $N \rightarrow \infty$ and,
by using the asymptotic random matrix theory of \cite{Tulino-TIT05}, we arrive at a computationally
efficient algorithm.
Let $\Upsilon^{(k)}_{i:A} = \lim_{N \rightarrow \infty} \MMSE^{(k)}_{i:A}$ and define
$\Gamma^{(k)}_{i:A} = 1/\Upsilon^{(k)}_{i:A} - 1$ and then a direct application of
\cite[Lemma 1]{Tulino-TIT05} yields $\Gamma^{(k)}_{i:A}$ as the solution of the fixed-point equation
\begin{equation} \label{eq:fixed-pt-eq}
\Gamma^{(k)}_{i:A} = \gamma Q_{\pi_k} \sum_{m=1}^{B/\gamma} \frac{\beta_{m,\pi_k}^2}
{1 + \sum_{j=i}^A \frac{\beta_{m,\pi_j}^2 Q_{\pi_j}}{1 + \Gamma^{(k)}_{i:A}}}.
\end{equation}
Combining (\ref{eq:fixed-pt-eq}) with the iterative algorithm \cite[Algorithm 1]{Tulino-TWC06}
that converges to the solution of (\ref{eq:KKT-finite-dim}), we obtain Algorithm \ref{alg:1}
below. (for notation simplicity, $\pi_k = k$ is assumed for all $k = 1,\ldots, A$).

\begin{algorithm}
\caption{Algorithm for weighted sum-rate maximization} \label{alg:1}
\begin{enumerate}
\item Initialize $\Qm(0) = \frac{Q}{A} \Id_A$.
\item For $\ell = 0,1,2,\cdots$, iterate until the following solution settles:
\begin{equation} \label{eq:algo3a}
Q_k(\ell+1) = Q \frac{\sum_{i=1}^k \Delta_i (1 - \Upsilon^{(k)}_{i:A}(\ell))}
           {\sum_{j=1}^A \sum_{i=1}^j \Delta_i ( 1 - \Upsilon^{(j)}_{i:A}(\ell))}
\end{equation}
for $k = 1, \ldots, A$, where $\Upsilon^{(k)}_{i:A}(\ell) = 1/(1 + \Gamma^{(k)}_{i:A}(\ell))$
and $\Gamma^{(k)}_{i:A}(\ell)$ is obtained as the solution (also obtained by iteration) of
(\ref{eq:fixed-pt-eq}) for powers $Q_k = Q_k(\ell)$.
\item Denote by $\Gamma^{(k)}_{i:A}(\infty)$, $\Upsilon^{(k)}_{i:A}(\infty)$, and
$Q_k(\infty)$ the fixed points reached by the iterations in step 2). If the condition
\[ Q \sum_{i=1}^k \Delta_i \Gamma^{(k)}_{i:A}(\infty) ~\leq ~
\sum_{j=1}^A \sum_{i=1}^j \Delta_i \left( 1 -\Upsilon^{(j)}_{i:A}(\infty) \right)\]
is satisfied for all $k$ such that $Q_k(\infty) = 0$, then stop. Otherwise, set
$Q_k = 0$ for $k$ corresponding to the lowest value of
$\sum_{i=1}^k \Delta_i \Gamma^{(k)}_{i:A}(\infty)$ and repeat steps 2) and 3).
\end{enumerate}
\end{algorithm}

After group powers $Q_k^\star = Q_k(\infty)$ have been obtained from Algorithm \ref{alg:1},
it remains to compute the ergodic group rates. We have
\begin{equation} \label{individual-rate}
R_{\pi_k} = \frac{1}{N} \EE \left[ \log \left| \Id + \Thetam_{k:A} \right| \right]
 - \frac{1}{N} \EE \left[ \log \left| \Id + \Thetam_{k+1:A} \right| \right].
\end{equation}
In the limit for $N\rightarrow \infty$, we can use the asymptotic analytical expression for the
mutual information given in \cite{Aktas-TIT06}. Adapting \cite[Result 1]{Aktas-TIT06} to our case,
we obtain $\lim_{N \rightarrow \infty} \frac{1}{N}\EE \left[ \log \left| \Id + \Thetam_{k:A}
\right| \right]$ from expression (21) in \cite{Moon-Caire-ICC10}.

\vspace{3mm}
\section{Introducing fairness} \label{sec:sche}

In a finite dimensional system, a dynamic scheduling policy allocates powers and rates in order to
obtain the system ergodic rate point (i.e. the time-average user rates) as close to the solution
of (\ref{eq:sche}) as possible. This can be systematically obtained by the stochastic
optimization approach of \cite{Georgiadis-Neely-Tassiulas-04, ShiraniMehr-Caire-Neely-submit09},
based on the idea of ``virtual queues''. For a deterministic network, we can exploit Lagrangian duality
with outer subgradient iteration, where the Lagrange dual variables play the roles of
the virtual queue backlogs of the dynamic scheduling. In the large system limit,
the channel uncertainty disappears and the multi-cell MU-MIMO system becomes indeed deterministic.
Hence, problem (\ref{eq:sche}) for large $N$ can be addressed directly, using Lagrangian duality.

We rewrite (\ref{eq:sche}) using auxiliary variables $\rv = [r_1, \cdots, r_A]$ as follows:
\begin{align} \label{eq:aux1}
\max_{\rv,\Qm,\pi} \;\;\; &  g(\rv) \nonumber \\
\mbox{s.t.} \;\;\; &  r_{\pi_k} \leq \frac{1}{N}\EE \left [ \log \frac{\left | \Id + \Thetam_{k:A}
\right |}{ \left | \Id + \Thetam_{k+1:A} \right |} \right ], \;\; \forall \; k, \nonumber \\
& \trace(\Qm) \leq Q.
\end{align}
The Lagrangian function of problem (\ref{eq:aux1}) is given by
\begin{align} \label{eq:lagrange}
& \Lc(\rv,\Qm,\pi,\muv) \nonumber \\
&= \underbrace{g(\rv) - \sum_{k=1}^A r_{\pi_k} \mu_{\pi_k}}_{f_{\muv}(\rv)}
+ \underbrace{\sum_{k=1}^A \mu_{\pi_k} \frac{1}{N} \EE \left[ \log \frac{\left| \Id + \Thetam_{k:A}
\right|}{\left| \Id + \Thetam_{k+1:A}\right|} \right]}_{h_{\muv}^N(\Qm,\pi)}
\end{align}
where $\muv$ denotes the dual variables for the rate constraints.
The Lagrange dual function for (\ref{eq:lagrange}) is given by
\begin{align} \label{eq:dual-ftn}
\Gc(\muv) \;=\; & \max_{\rv,\Qm,\pi} \;\;\; \Lc(\rv,\Qm,\pi,\muv) \nonumber \\
\;=\; & \underbrace{\max_{\rv} \;\;\; f_{\muv}(\rv)}_{(a)}
  + \underbrace{\max_{\Qm,\pi} \;\;\; h_{\muv}^N(\Qm,\pi)}_{(b)}
\end{align}
and it is obtained by the decoupled maximization in (a) (with respect to $\rv$) and in (b)
(with respect to $\Qm, \pi$) in (\ref{eq:dual-ftn}). Finally, we solve the dual problem defined as
\begin{equation} \label{eq:dual-prob}
\min_{\muv \geq \zerov} \;\;\; \Gc(\muv).
\end{equation}
Since the maximization of (b) is a weighted ergodic sum-rate maximization for weights $\Wm = \muv$,
it follows that the optimal $\pi^*$ is the permutation that sorts $\mu$ in non-decreasing order
(see Section \ref{sec:wesrm}). For $\pi = \pi^*$, the Lagrangian function is concave in $\rv$ and
$\Qm$ and convex in $\mu$ and the solution (saddle point of the min--max problem) can be found via
inner--outer iterations.

{\bf Inner Problem}: For given $\muv$, we solve the maximization problem in (\ref{eq:dual-ftn})
with respect to $\rv$, $\Qm$ and $\pi$.

$\bullet$ Subproblem (a): Since $f_{\muv}(\rv)$ is concave in $\rv \geq 0$, the maximum of
$f_{\muv}(\rv)$ is achieved by imposing the KKT conditions:
\begin{align} \label{eq:sola}
\frac{\partial g(\rv)}{\partial r_k} - \mu_k \leq 0, \;\; \forall k
\end{align}
where the equality must hold for all $k$ such that the solution is positive, i.e., $r^*_k > 0$.

$\bullet$ Subproblem (b): As said, this problem is a weighed ergodic sum-rate maximization and
can be solved for $N \rightarrow \infty$ by Algorithm \ref{alg:1}.

{\bf Outer Problem}: Once the optimal $\rv^*$, $\Qm^*$ and $\pi^*$ are obtained for given $\muv$,
the minimization of $\Gc(\muv)$ with respect to $\muv \geq \zerov$ can be performed by a
subgradient-based method. For any fixed $\muv'$ and $\muv$, we have
\begin{align} \label{eq:subgrad}
\Gc(\muv') & =\; \max_{\rv} \; f_{\muv'}(\rv) + \max_{\Qm} \; h_{\muv'}^\infty(\Qm,\pi^*) \nonumber \\
&\geq f_{\muv'}(\rv^*) + h_{\muv'}^N(\Qm^*,\pi^*) \nonumber \\
&= \Gc(\muv) + \sum_{k=1}^A \left( \mu'_{\pi_k^*} - \mu_{\pi_k^*} \right) \left( R_{\pi_k^*}
 - r_{\pi_k^*} \right)
\end{align}
where $R_{\pi_k^*}$ is the rate of user group $\pi_k^*$ obtained from Algorithm \ref{alg:1} with weights $\muv$.
Then, the vector with components $R_{\pi^*_k} - r_{\pi^*_k}$ is the subgradient for $\Gc(\muv)$ and
dual variable $\mu_{\pi_k}$ is updated at each outer iteration $n$ as
\begin{equation} \label{eq:mu-update}
\mu_{\pi^*_k}(n+1) = \mu_{\pi^*_k}(n) - s(n) \left( R_{\pi^*_k}(n) - r_{\pi^*_k}(n) \right),
 \;\;\; \forall \; k
\end{equation}
for some step size $s(n)>0$ which can be determined efficiently by a back-tracking line search method \cite{Boyd-Vandenberghe-04}. It should be noticed that by setting $s(n)=1$ this subgradient update plays
the role of the virtual queue update in the dynamic scheduling policy of
\cite{Georgiadis-Neely-Tassiulas-04, ShiraniMehr-Caire-Neely-submit09}.

As an application example of the above general optimization, we focus on the two special cases,
{\em proportional fairness scheduling} (PFS) and {\em hard fairness scheduling} (HFS), also known as
max-min fairness scheduling.
PFS corresponds to the network utility function $g(\rv) = \sum_{k=1}^K C \log (r_k)$ for some constant
$C>0$ and the optimality condition (\ref{eq:sola}) yields $r_k (n) = C/\mu_k (n)$ at outer
iteration $n$.

In the case of HFS, the network utility function is given as $g(\rv) = \min_{k=1,\cdots,A} C r_k$ and
the corresponding optimization problem can be rewritten as
\begin{align} \label{eq:hfs}
\max_{r,\Qm,\pi} \;\;\; & C r \nonumber \\
\mbox{s.t.} \;\;\; & r \leq \frac{1}{N}\EE \left [ \log \frac{\left | \Id + \Thetam_{k:A} \right |}
 { \left | \Id + \Thetam_{k+1:A} \right |} \right ], \; \forall k \nonumber \\
& \trace(\Qm) \leq Q.
\end{align}
Replicating the former approach for this problem, with the corresponding changes, we find
$r(n) = \frac{1}{A} \sum_{k=1}^A R_{\pi_k}(n)$ at outer iteration $n$.

\vspace{2mm}
\section{Numerical results and discussion} \label{sec:result}

\begin{figure}
\centering
\subfigure[Proportional fairness scheduling]{
\includegraphics[width=3.5in]{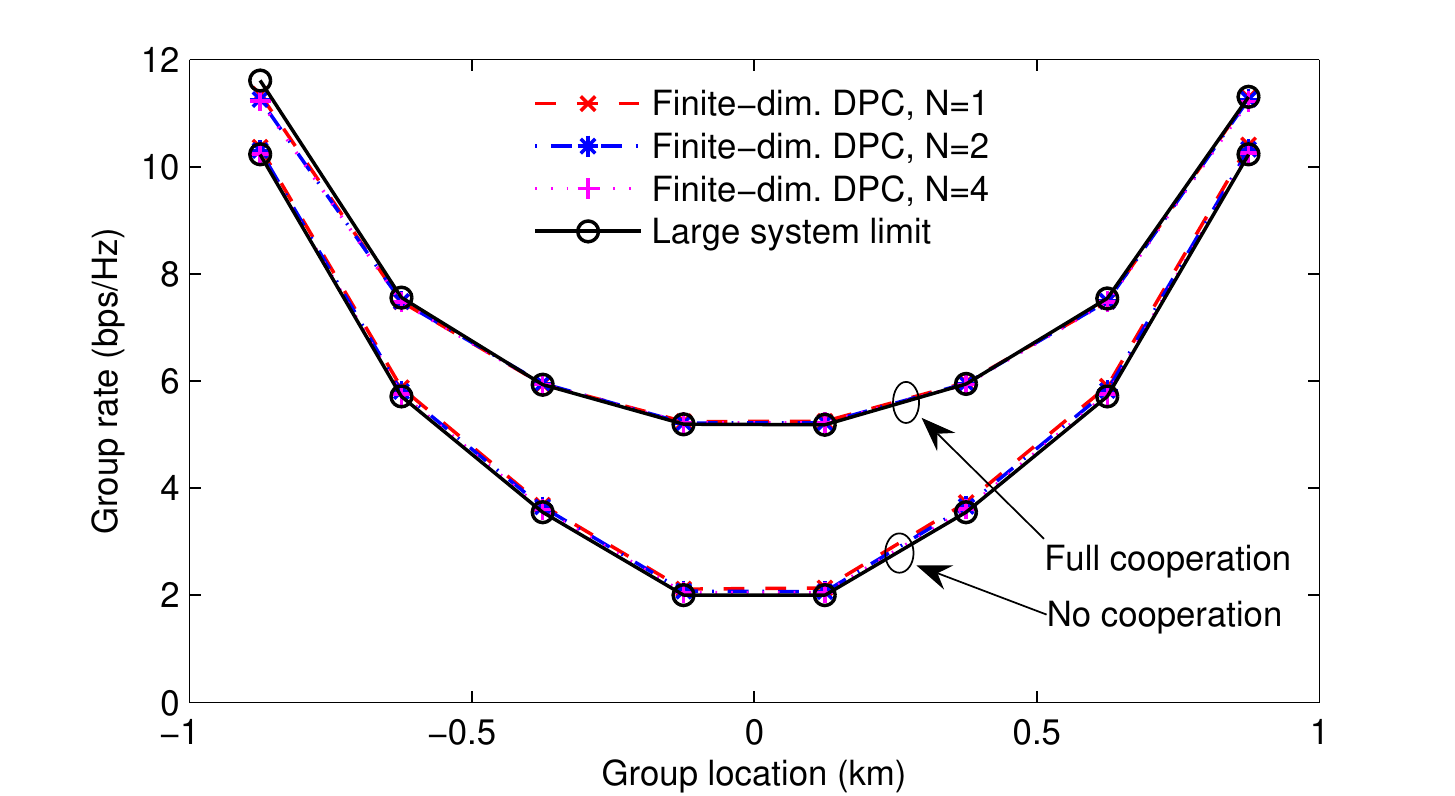}
\label{subfig:2cellpfs}
}
\subfigure[Hard fairness scheduling]{
\includegraphics[width=3.5in]{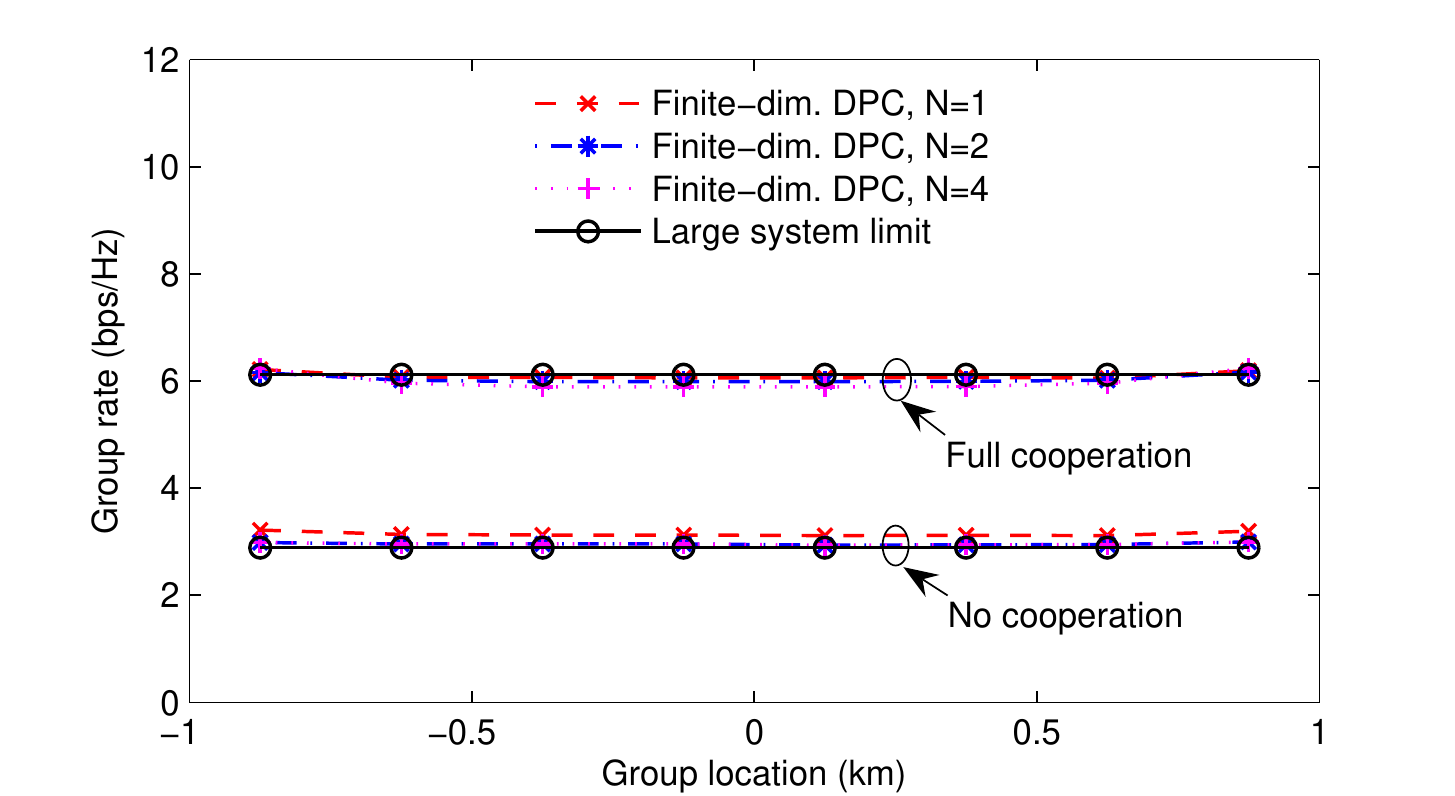}
\label{subfig:2cellhfs}
}
\caption{Individual group rates under (a) PFS and (b) HFS for $\gamma=4$ and $K=8$ in the 2-cell model}
\label{fig:2cell}
\end{figure}

We present numerical examples for a one-dimensional 2-cell model ($M=2$) and two-dimensional
three-sectored 7-cell model ($M=21$). In both models, the system parameters and pathloss model are
based on the mobile WiMAX system evaluation specification \cite{Wimax-eval06}, except for cell radius
1.0 km and no shadow fading assumption.
In the 2-cell model, we assume two one-sided cells with BSs located at -1 km and 1 km with $\gamma=4$,
and $K=8$ user groups equally spaced between the two BSs. We consider the case of full BS cooperation
and no cooperation with a symmetric partition of user groups, namely: $L=2$, $\Kc_1=\{1,2,3,4\}$ and
$\Kc_2=\{5,6,7,8\}$.
Fig. \ref{fig:2cell} illustrates the individual group rates (i.e. mean user rates of each group)
in the large system limit as a function of group locations and compares them with the achievable rates
obtained by Monte Carlo simulation in finite dimensions with $N$ = 1, 2, or 4.
In the simulation with finite $N$, the BSs are equipped with $\gamma N$ antennas and
$N$ users are located at each of $K$ locations. Channel vectors are randomly generated and the
dynamic scheduling \cite{ShiraniMehr-Caire-Neely-submit09} and DPC precoding with water-filling algorithm
\cite{Yu-TIT06} are applied to each realization of channel vectors.
In plots (a) and (b), the PFS and HFS are considered, respectively. Remarkably, the rates obtained by
the large system analysis almost overlap with those obtained by the finite-dimensional simulations,
even for $N=1$. Notice that the dynamic scheduling policy should provide multiuser diversity gain
and achieve in general higher rates than the large system limit which is not able to exploit
the dynamic fluctuations of the small-scale fading due to ``channel hardening'' effect. However,
it appears that in the regime where the pathloss is dominant over the randomness of multi-antenna
channels and the number of users is not much larger than the number of BS antennas, the multiuser
diversity gain is negligible and the asymptotic analysis produces rate points very close to the outputs
of simulations with dynamic scheduling. Notice that in HFS case all the users achieve the same
individual rate which is slightly higher than the smallest rate under the PFS.

\begin{figure}
\centering
\subfigure[No cooperation]{
\includegraphics[width=2.9in]{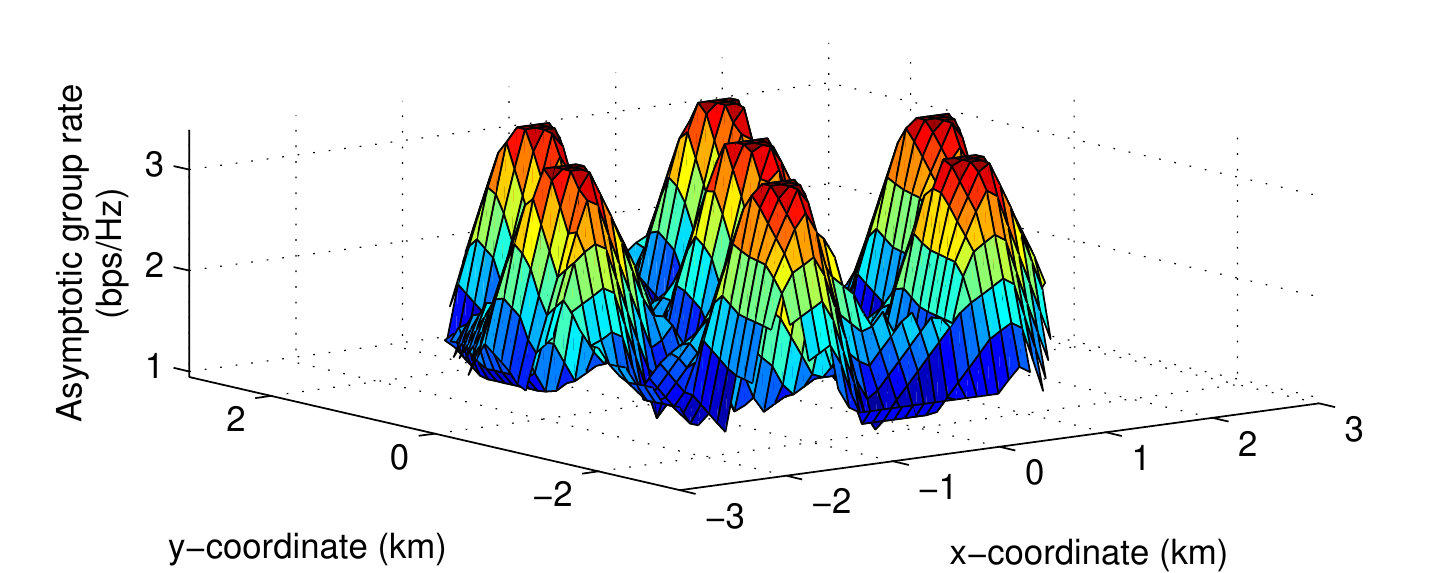}
\label{fig:nocoop}
}
\subfigure[Cooperation among co-located sectors]{
\includegraphics[width=2.9in]{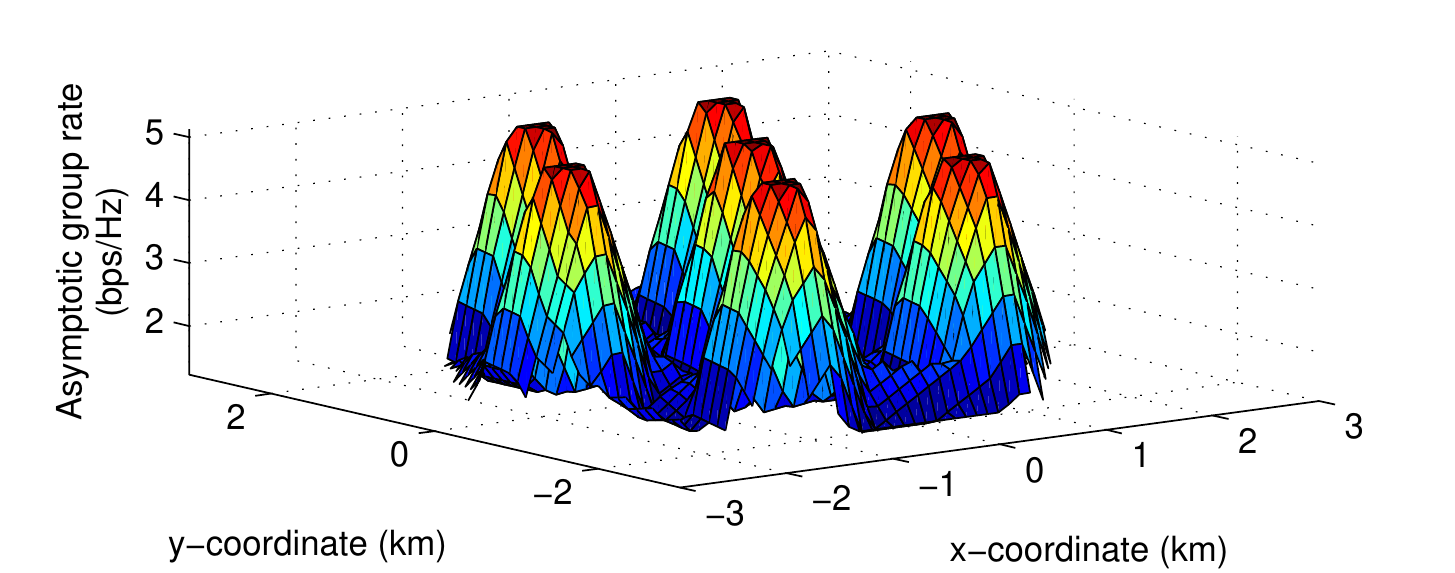}
\label{fig:sectcoop}
}
\subfigure[Full cooperation over 7 cells]{
\includegraphics[width=2.9in]{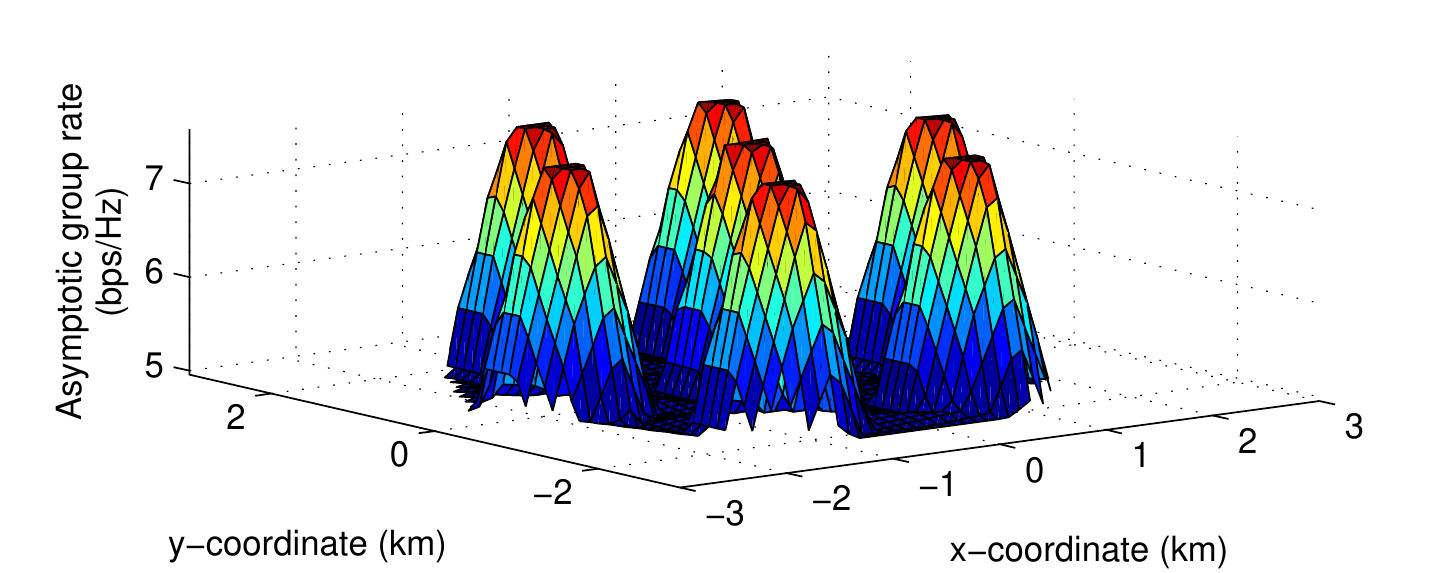}
\label{fig:bscoop}
}
\caption{Ergodic group rate distribution under PFS in the 7-cell model}
\label{fig:2dsim}
\end{figure}

Using the proposed asymptotic analysis, validated with the simple 2-cell model, we can obtain
ergodic rate distributions for much larger systems, for which a full-scale simulation would be very
demanding. We consider a two-dimensional cell layout where 7 hexagonal cells form a network and
each cell consists of three 120-degree sectors. Three BSs are co-located at the center of each cell
such that each BS handles one sector in no cooperation case. Each sector is split into 4 diamond-shaped
equal-area grids and one user group is placed at the center of each grid. Therefore there are total
$M=21$ BSs and $K=84$ user groups in the network. In addition, we assume the 7 cells are wrap-around
in a torus topology, such that each cell is virtually surrounded by the other 6 cells and all
the cells have the symmetric inter-cell interference distribution. The antenna orientation and pattern
follows \cite{IEEE80216m-EMD09} and the non-ideal antenna pattern generates inter-cell interference
even between co-located sectors with no cooperation. This model indeed conjectures very large
scale of cellular networks in that the dominant inter-cell interference and effective inter-cell
cooperation gain comes from directly neighboring cells.
Fig. \ref{fig:2dsim} shows the user rate distribution under three levels of cooperation,
(a) no cooperation ($L=21$), (b) cooperation among co-located 3 sectors ($L=7$), and
(c) full cooperation over 7-cell network ($L=1$). The simulation is very intricate especially in case (c).
From the asymptotic rate results, it is shown that in case (b), the cooperation gain over case (a) is
primarily obtained for users around cell centers, whereas the gain is achieved over all the
locations in case (c).

\bibliography{isit10-mimo-multicell-asympt-cr}

\begin{thebibliography}{10}
\providecommand{\url}[1]{#1}
\csname url@samestyle\endcsname
\providecommand{\newblock}{\relax}
\providecommand{\bibinfo}[2]{#2}
\providecommand{\BIBentrySTDinterwordspacing}{\spaceskip=0pt\relax}
\providecommand{\BIBentryALTinterwordstretchfactor}{4}
\providecommand{\BIBentryALTinterwordspacing}{\spaceskip=\fontdimen2\font plus
\BIBentryALTinterwordstretchfactor\fontdimen3\font minus
  \fontdimen4\font\relax}
\providecommand{\BIBforeignlanguage}[2]{{%
\expandafter\ifx\csname l@#1\endcsname\relax
\typeout{** WARNING: IEEEtran.bst: No hyphenation pattern has been}%
\typeout{** loaded for the language `#1'. Using the pattern for}%
\typeout{** the default language instead.}%
\else
\language=\csname l@#1\endcsname
\fi
#2}}
\providecommand{\BIBdecl}{\relax}
\BIBdecl

\bibitem{Caire-Shamai-TIT03}
G.~Caire and S.~{Shamai (Shitz)}, ``{On the achievable throughput of a
  multiantenna Gaussian broadcast channel},'' \emph{{IEEE} Trans. on Inform.
  Theory}, vol.~49, pp. 1691--1706, July 2003.

\bibitem{Goldsmith-JSAC03}
A.~J. Goldsmith, S.~A. Jafar, N.~J. Jindal, and S.~Vishwanath, ``{Capacity
  limits of MIMO channels},'' \emph{{IEEE} J. Select. Areas Commun.}, vol.~21,
  pp. 684--702, June 2003.

\bibitem{Weingarten-Steinberg-Shamai-TIT06}
H.~Weingarten, Y.~Steinberg, and S.~{Shamai (Shitz)}, ``{The capacity region of
  the Gaussian multiple-input multiple-output broadcast channel},''
  \emph{{IEEE} Trans. on Inform. Theory}, vol.~52, pp. 3936--3964, Sept. 2006.

\bibitem{Wyner-TIT94}
A.~D. Wyner, ``{Shannon-theoretic approach to a Gaussian cellular multiple
  access channel},'' \emph{{IEEE} Trans. on Inform. Theory}, vol.~40, pp.
  1713--1727, Nov. 1994.

\bibitem{Shamai-Wyner-TIT97}
S.~{Shamai (Shitz)} and A.~D. Wyner, ``{Information-theoretic considerations
  for symmetric, cellular, multiple-access fading channels -- Part I \& II},''
  \emph{{IEEE} Trans. on Inform. Theory}, vol.~43, pp. 1877--1894, Nov. 1997.

\bibitem{Somekh-Shamai-TIT00}
O.~Somekh and S.~{Shamai (Shitz)}, ``{Shannon-theoretic approach to a Gaussian
  cellular multiple-access channel with fading},'' \emph{{IEEE} Trans. on
  Inform. Theory}, vol.~46, pp. 1401--1425, July 2000.

\bibitem{Sanderovich-TIT09}
A.~Sanderovich, O.~Somekh, H.~Poor, and S.~{Shamai (Shitz)}, ``{Uplink macro
  diversity of limited backhaul cellular network},'' \emph{{IEEE} Trans. on
  Inform. Theory}, vol.~55, pp. 3457--3478, Aug. 2009.

\bibitem{Wimax-eval06}
{WiMAX Forum}, ``{Mobile WiMAX -- Part I: A technical overview and performance
  evaluation},'' Tech. Rep., Aug. 2006.

\bibitem{Viswanath-Tse-Laroia-TIT02}
P.~Viswanath, D.~Tse, and R.~Laroia, ``{Opportunistic beamforming using dumb
  antennas},'' \emph{{IEEE} Trans. on Inform. Theory}, vol.~48, pp. 1277--1294,
  June 2002.

\bibitem{Georgiadis-Neely-Tassiulas-04}
L.~Georgiadis, M.~Neely, and L.~Tassiulas, \emph{{Resource Allocation and
  Cross-Layer Control in Wireless Networks}}.\hskip 1em plus 0.5em minus
  0.4em\relax Foundations and Trends in Networking, 2006, vol.~1, no.~1.

\bibitem{ShiraniMehr-Caire-Neely-submit09}
H.~Shirani-Mehr, G.~Caire, and M.~J. Neely, ``{MIMO downlink scheduling with
  non-perfect channel state knowledge},'' \emph{submitted to IEEE Trans. on
  Commun. (posted on arXiv:0904.1409 [cs.IT])}, 2009.

\bibitem{Mo-TNET00}
J.~Mo and J.~Walrand, ``{Fair end-to-end window-based congestion control},''
  \emph{{IEEE/ACM} Trans. on Networking}, vol.~8, pp. 556--567, Oct. 2000.

\bibitem{Tulino-TIT05}
A.~M. Tulino, A.~Lozano, and S.~Verdu, ``{Impact of antenna correlation on the
  capacity of multiantenna channels},'' \emph{{IEEE} Trans. on Inform. Theory},
  vol.~7, pp. 2491--2509, July 2005.

\bibitem{Aktas-TIT06}
D.~Aktas, M.~N. Bacha, J.~S. Evans, and S.~V. Hanly, ``{Scaling results on the
  sum capacity of cellular networks with MIMO links},'' \emph{{IEEE} Trans. on
  Inform. Theory}, vol.~52, pp. 3264--3274, July 2006.

\bibitem{Moon-Caire-ICC10}
S.-H. Moon, H.~Huh, Y.-T. Kim, I.~Lee, and G.~Caire, ``{Weighted sum-rate of
  multi-cell MIMO downlink channels in the large system limit},'' in
  \emph{Proc. {IEEE} Int. Conf. on Commun. (ICC)}, Cape Town, South Africa, May
  2010.

\bibitem{Huh-Tulino-Caire-CISS10}
H.~Huh, A.~M. Tulino, and G.~Caire, ``{Network MIMO large-system analysis and
  the impact of CSIT estimation},'' in \emph{Proc. Conf. on Inform. Sciences
  and Systems (CISS)}, Princeton, NJ, March 2010.

\bibitem{Jindal-TIT03}
N.~Jindal and A.~J. Goldsmith, ``{Capacity and optimal power allocation for
  fading broadcast channels with minimum rates},'' \emph{{IEEE} Trans. on
  Inform. Theory}, vol.~49, pp. 2895--2909, November 2003.

\bibitem{Jindal-TIT05}
N.~Jindal, W.~Rhee, S.~Vishwanath, S.~A. Jafar, and A.~J. Goldsmith, ``{Sum
  power iterative water-filling for multi-antenna Gaussian broadcast
  channels},'' \emph{{IEEE} Trans. on Inform. Theory}, vol.~51, pp. 1570--1580,
  April 2005.

\bibitem{Tse-05}
D.~Tse and P.~Viswanath, \emph{{Fundamentals of Wireless Communication}}.\hskip
  1em plus 0.5em minus 0.4em\relax Cambridge University Press, 2005.

\bibitem{Tulino-TWC06}
A.~M. Tulino, A.~Lozano, and S.~Verdu, ``{Capacity-achieving input covariance
  for single-user multi-antenna channels},'' \emph{{IEEE} Trans. on Wireless
  Commun.}, vol.~5, pp. 662--671, March 2006.

\bibitem{Boyd-Vandenberghe-04}
S.~Boyd and L.~Vandenberghe, \emph{{Convex Optimization}}.\hskip 1em plus 0.5em
  minus 0.4em\relax Cambridge University Press, 2004.

\bibitem{Yu-TIT06}
W.~Yu, ``{Sum-capacity computation for the Gaussian vector broadcast channel
  via dual decomposition},'' \emph{{IEEE} Trans. on Inform. Theory}, vol.~52,
  pp. 754--759, Feb. 2006.

\bibitem{IEEE80216m-EMD09}
{IEEE 802.16 broadband wireless access working group}, ``{IEEE 802.16m
  evaluation methodology document (EMD)},'' Tech. Rep., Jan. 2009.

\end{thebibliography}

\end{document}